\documentclass[10pt, conference]{IEEEtran}
\IEEEoverridecommandlockouts
\usepackage{cite}
\usepackage{amsmath,amssymb,amsfonts}
\usepackage{algorithmic}
\usepackage{graphicx}
\usepackage{textcomp}
\usepackage{hyperref}
\hypersetup{
  colorlinks   = true, 
  urlcolor     = blue, 
  linkcolor    = blue, 
  citecolor   = blue 
}

\def\BibTeX{{\rm B\kern-.05em{\sc i\kern-.025em b}\kern-.08em
    T\kern-.1667em\lower.7ex\hbox{E}\kern-.125emX}}
    
\usepackage{xcolor}

\begin{document}

\title{Learning Quantum Phase Estimation by Variational Quantum Circuits}





\author{
\IEEEauthorblockN{
    Chen-Yu Liu \IEEEauthorrefmark{1}\IEEEauthorrefmark{2}\IEEEauthorrefmark{5}, Chu-Hsuan Abraham Lin \IEEEauthorrefmark{1}\IEEEauthorrefmark{3}, and
    Kuan-Cheng Chen
    \IEEEauthorrefmark{4}
}
\IEEEauthorblockA{\IEEEauthorrefmark{1} Hon Hai (Foxconn) Research Institute, Taipei, Taiwan}
\IEEEauthorblockA{\IEEEauthorrefmark{2} Graduate Institute of Applied Physics, National Taiwan University, Taipei, Taiwan}
\IEEEauthorblockA{\IEEEauthorrefmark{3} Department of Electrical and Electronic Engineering, Imperial College London, London, UK}
\IEEEauthorblockA{\IEEEauthorrefmark{4} Centre for Quantum Engineering, Science and Technology (QuEST), Imperial College London, London, UK}

\IEEEauthorblockA{Email:\IEEEauthorrefmark{5}  d10245003@g.ntu.edu.tw}
}

\date{\today}
\maketitle

\begin{abstract}
Quantum Phase Estimation (QPE) stands as a pivotal quantum computing subroutine that necessitates an inverse Quantum Fourier Transform (QFT). However, it is imperative to recognize that enhancing the precision of the estimation inevitably results in a significantly deeper circuit. We developed a variational quantum circuit (VQC) approximation to reduce the depth of the QPE circuit, yielding enhanced performance in noisy simulations and real hardware. Our experiments demonstrated that the VQC outperformed both Noisy QPE simulation and standard QPE on real hardware by reducing circuit noise. This VQC integration into quantum compilers as an intermediate step between input and transpiled circuits holds significant promise for quantum algorithms with deep circuits. Future research will explore its potential applicability across various quantum computing hardware architectures.
\end{abstract}

\section{Introduction}


Quantum computing, grounded in the principles of quantum mechanics~\cite{qc1, qc2}, presents a transformative approach to calculations and problem-solving schemes. Its potential to solve specific problems at significantly enhanced speeds compared to classical computing holds promise for diverse applications. However, the realization of fault-tolerant quantum computing, where quantum algorithms can assert clear advantages over classical counterparts, remains a prospect for the future, as we presently reside within the confines of the Noisy Intermediate-scale Quantum (NISQ) era~\cite{qc2, nisq1, nisq2}. During this phase, the inherent errors within quantum systems impose constraints on the number of consecutive quantum operations, commonly referred to as circuit depth.


Quantum Phase Estimation (QPE) is a fundamental quantum algorithm that plays a crucial role in various quantum computing applications. It is designed to determine the phase factor of a unitary operator. QPE is a core component in many quantum algorithms, such as Shor's algorithm \cite{qc0} for integer factorization and quantum algorithms for solving linear equations \cite{hhl}. QPE provides a powerful means of extracting valuable information from quantum states, enabling quantum computers to address complex problems more efficiently than their classical counterparts. By accurately estimating phase information, QPE contributes to the advancement of quantum computing's capabilities, making it a key area of research and development in the field of quantum algorithms.

However, in the context of approximating a phase with a precision of $2^{-n}$ and achieving a success probability of at least $1-\epsilon$, a substantial computational effort is required, encompassing approximately $O(t^2)$ operations and utilizing a quantum register of size $t$ qubits, as outlined by Neilson and Chuang \cite{qc0}, where $t$ is defined as 
\begin{equation}
    t = n + \lceil \log(2+\frac{1}{2\epsilon}) \rceil.
\end{equation}
Notably, during the NISQ era, the circuit depth of the quantum circuit, proportional to $O(t^2)$, poses significant challenges in preserving the integrity of computational results obtained on real quantum devices. As a consequence, strategies to mitigate errors and maintain the fidelity of quantum circuit information have become the main concerns within this domain.


Variational Quantum Circuits (VQCs) have a diverse range of applications, including Quantum Neural Networks (QNNs) \cite{qml1, qml2, qml3}, Quantum Generative Adversarial Networks (QGANs) \cite{qgan1, qgan2, qgan3, qgan4, qgan5}, Quantum Reinforcement Learning (QRL) \cite{qrl1, qrl2}, as well as the capability to address the ground state problems of Ising Hamiltonians \cite{toising1, toising2}. This latter capability, in particular, opens the door to wide practical applications, extending to the combinatorial optimization problems, and the simulation of quantum chemistry \cite{vqe0, vqe1, vqe2, vqe3, clustervqe1, lssa, fmrs, pqs, rlqls}. 


In this work, we present a novel approach aimed at training a VQC to replicate the outcomes achieved through QPE. Our methodology leads to a significant reduction in circuit depth to the order of $O(pn)$ for VQCs employing linear entanglement, where $p$ corresponds to the number of layers within the associated VQC. We present results obtained from both classical simulations and practical quantum hardware implementations. In Section~\ref{sec:learnQPE}, we introduce the specific QPE example employed in this study and delineate a VQC learning framework devised to capture the outcomes of the corresponding QPE. Subsequently, in Section~\ref{sec:result}, we present comprehensive findings derived from both simulation and quantum hardware implementations, following the aforementioned learning approach. Finally, in Section~\ref{sec:discuss}, we examined the observed data patterns, providing insight into the implications of our results, and conclude by outlining prospective directions for future research in this domain.

    


\section{Learning Quantum Phase Estimation}
\label{sec:learnQPE}
In this section, we first describe the quantum phase estimation example that we will use throughout this work and show how the corresponding circuit results can be learned using variational quantum circuits.
\subsection{Quantum Phase Estimation}
\label{sec:quantum_phase_estimation}

\begin{figure*}[ht]
\centering
\includegraphics[scale=0.32]{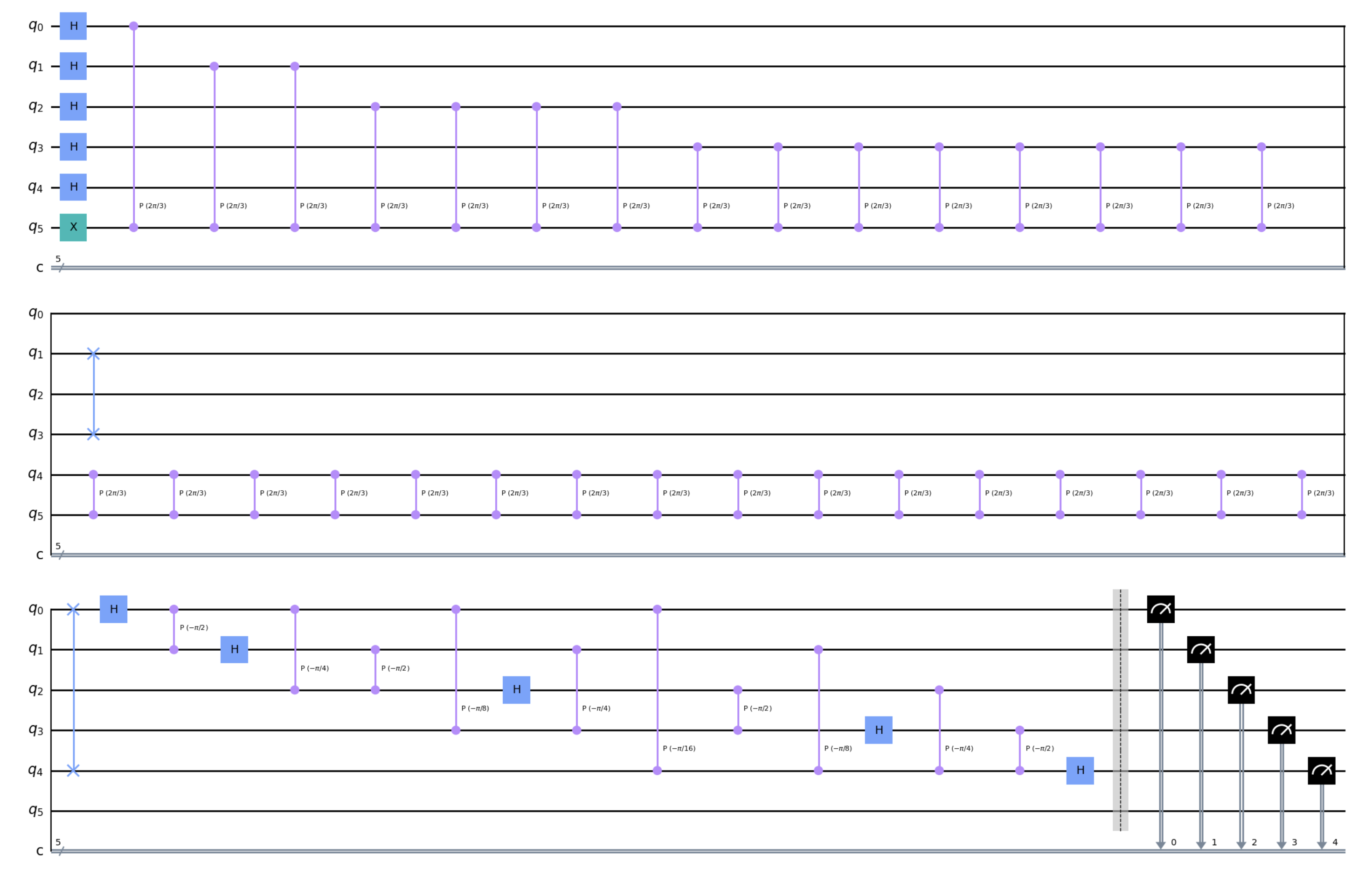}
\caption{The circuit for the QPE example in this work consists of 1 storage qubit and 5 counting qubits.}
\label{fig:qpe_circuit}
\end{figure*}

\begin{figure*}[ht]
\centering
\includegraphics[scale=0.32]{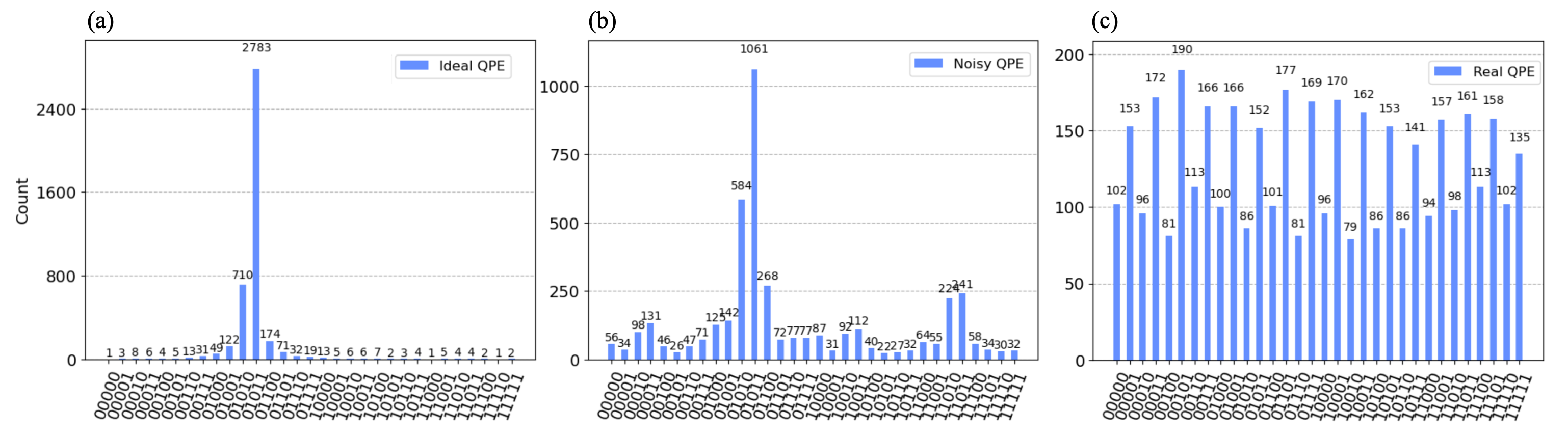}
\caption{Measurement results of the QPE circuit in Fig.~\ref{fig:qpe_circuit}. (a) Executed on an ideal simulator. (b) Executed on a noisy simulator using a noise model obtained from \textsf{ibm\_oslo} provided by IBM Quantum. (c) Executed on the 7-qubit IBM quantum computer, \textsf{ibm\_lagos}. The number of measurement shots for all cases is $N_{\text{shot}} = 4096$.}
\label{fig:qpe_result}
\end{figure*}

The QPE algorithm is employed to estimate the phase $\theta$ of the unitary operator $U | \psi \rangle = e^{2 \pi i \theta} | \psi \rangle$, where $| \psi \rangle$ represents the eigenvector of $U$ and $e^{2 \pi i \theta}$ corresponds to the eigenvalue. In this study, we illustrate the QPE algorithm using a specific case where the unitary operator $U$ is defined as:

\begin{equation}
U =
\begin{bmatrix}
1 & 0 \\
0 & e^{\frac{2}{3}i \pi}
\end{bmatrix}
\end{equation}
Here, the eigenvector $| \psi \rangle$ is equivalent to $|1\rangle = \begin{bmatrix}
0 \\
1
\end{bmatrix}$, and the corresponding eigenvalue is $e^{\frac{2}{3}i \pi}$. Consequently, the phase $\theta$ we aim to estimate through QPE is $\theta = \frac{1}{3}$. Storing the eigenstate for the operator $U$ necessitates 1 qubit, and we employ an additional 5 qubits for counting purposes. Thus, the complete circuit comprises 6 qubits, as depicted in Fig.~\ref{fig:qpe_circuit}. The measurement results for the QPE, obtained through ideal simulation, noisy simulation, and real hardware, are presented in Fig.~\ref{fig:qpe_result}, with a total of $N_{\text{shot}} = 4096$ measurement shots. The noisy simulation incorporates the noise model derived from \textsf{ibm\_lagos}, provided by IBM Quantum. The real hardware results are obtained from the 7-qubit IBM quantum computer, \textsf{ibm\_lagos}. In both the ideal and noisy simulations, the bit string with the highest measured probability is ``01011," which corresponds to ``11" in decimal. Consequently, the resulting value of $\theta$ for this circuit is $\theta = 11/2^5 = 0.344$.

\subsection{VQC for QPE}
\label{sec:vqc_for_QPE}

\begin{figure*}[ht]
\centering
\includegraphics[scale=0.26]{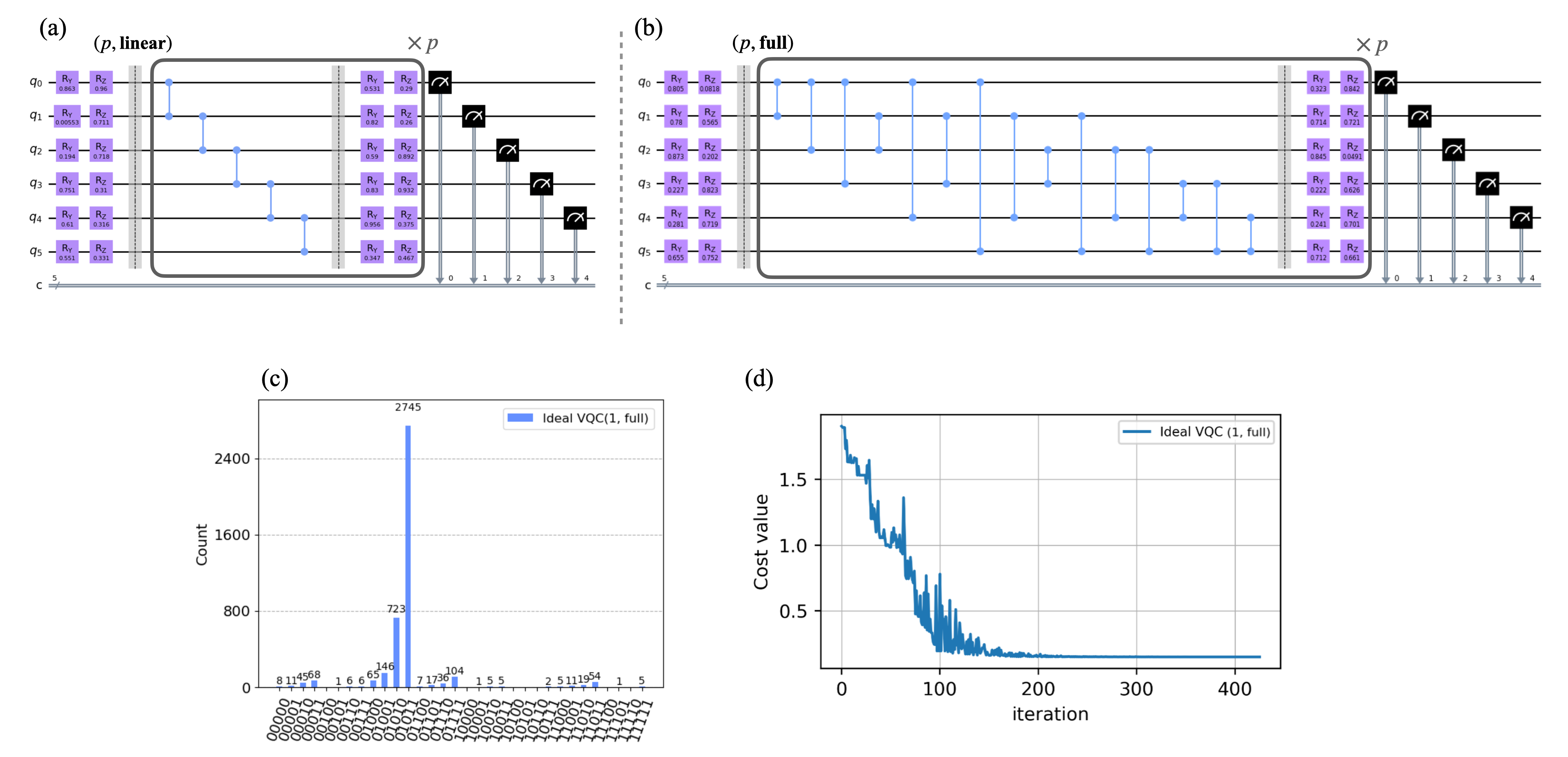}
\caption{(a) Circuit ansatz with a linear entangler and $p$ layers. (b) Circuit ansatz with a full entangler and $p$ layers. (c) Measurement results of the trained VQC under ideal simulation, with $p=1$, full entangler, and $N_{\text{shot}} = 4096$. (d) Cost values during the training under ideal simulation, with $p=1$, full entangler, and $N_{\text{shot}} = 4096$.}
\label{fig:vqc_ansatz_ideal_result}
\end{figure*}

VQC, equipped with trainable parameters, can be utilized for the preparation of specific quantum states \cite{sp1, sp2}. To implement a VQC, several essential components come into play, including an ansatz for the parameterized quantum circuit, a cost function, and a classical optimizer. The ansatz comprises adjustable rotational gates with a specified number of layers. As illustrated in Fig.~\ref{fig:vqc_ansatz_ideal_result}(a) and Fig.~\ref{fig:vqc_ansatz_ideal_result}(b), two types of entanglers, linear and full, are employed, each with a parameter $p$. It's noteworthy that in the linear case, the required circuit depth is $O(pn)$, while in the full case, it becomes $O(pn^2)$, where $n$ represents the number of qubits. Within each layer of rotational gates, only $R_y$ and $R_z$ gates are utilized. In Fig.~\ref{fig:vqc_ansatz_ideal_result}(a) and Fig.~\ref{fig:vqc_ansatz_ideal_result}(b), random angles fill the rotational gates for demonstration purposes.

To train a VQC with the target measurement results mirroring the QPE circuit, a dedicated cost function is established as follows:
\begin{equation}
\label{eq:cost}
\text{Cost} = \sum_{i=1}^{2^N} \sqrt{[ P(| \phi_i \rangle) - P^{\text{gt}}(| \phi_i \rangle) ]^2},
\end{equation}
Here, $N = 5$, representing the number of counting qubits in the QPE, and $P(|\phi_i \rangle)$ stands for the probability of measuring the computational basis $|\phi_i \rangle$. Meanwhile, $P^{\text{gt}}(|\phi_i \rangle)$ embodies the ground truth of the corresponding probability, making it a supervised learning task, thus necessitating the desired prediction outcome. The potential use cases of this proposed method will be discussed in Section~\ref{sec:discuss}.

Once the circuit ansatz is determined, and the cost function is defined, the ansatz parameters are tuned using a classical optimizer. In this study, the COBYLA optimizer is employed for training the VQC. One notable advantage of COBYLA is its efficiency, as it requires only one circuit execution per iteration, making it particularly advantageous for practical use cases.

\begin{figure*}[ht]
\centering
\includegraphics[scale=0.27]{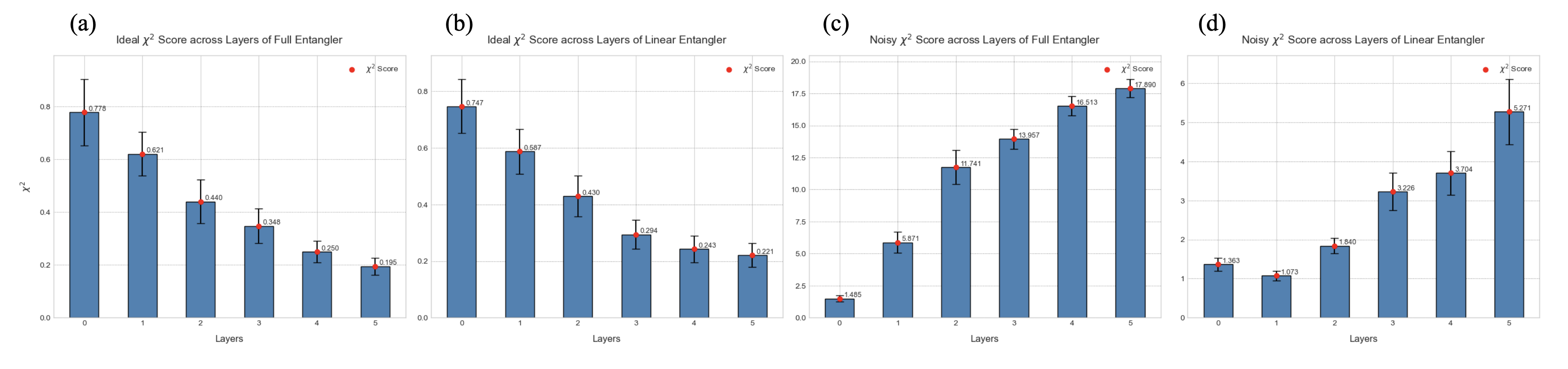}
\caption{$\chi^2$ score for different settings: (a) Ideal simulation with full entangler. (b) Ideal simulation with linear entangler. (c) Noisy simulation with full entangler. (d) Noisy simulation with linear entangler. }
\label{fig:vqc_layers_entanglers_benchmarking}
\end{figure*}

\begin{figure}[ht]
\centering
\includegraphics[scale=0.23]{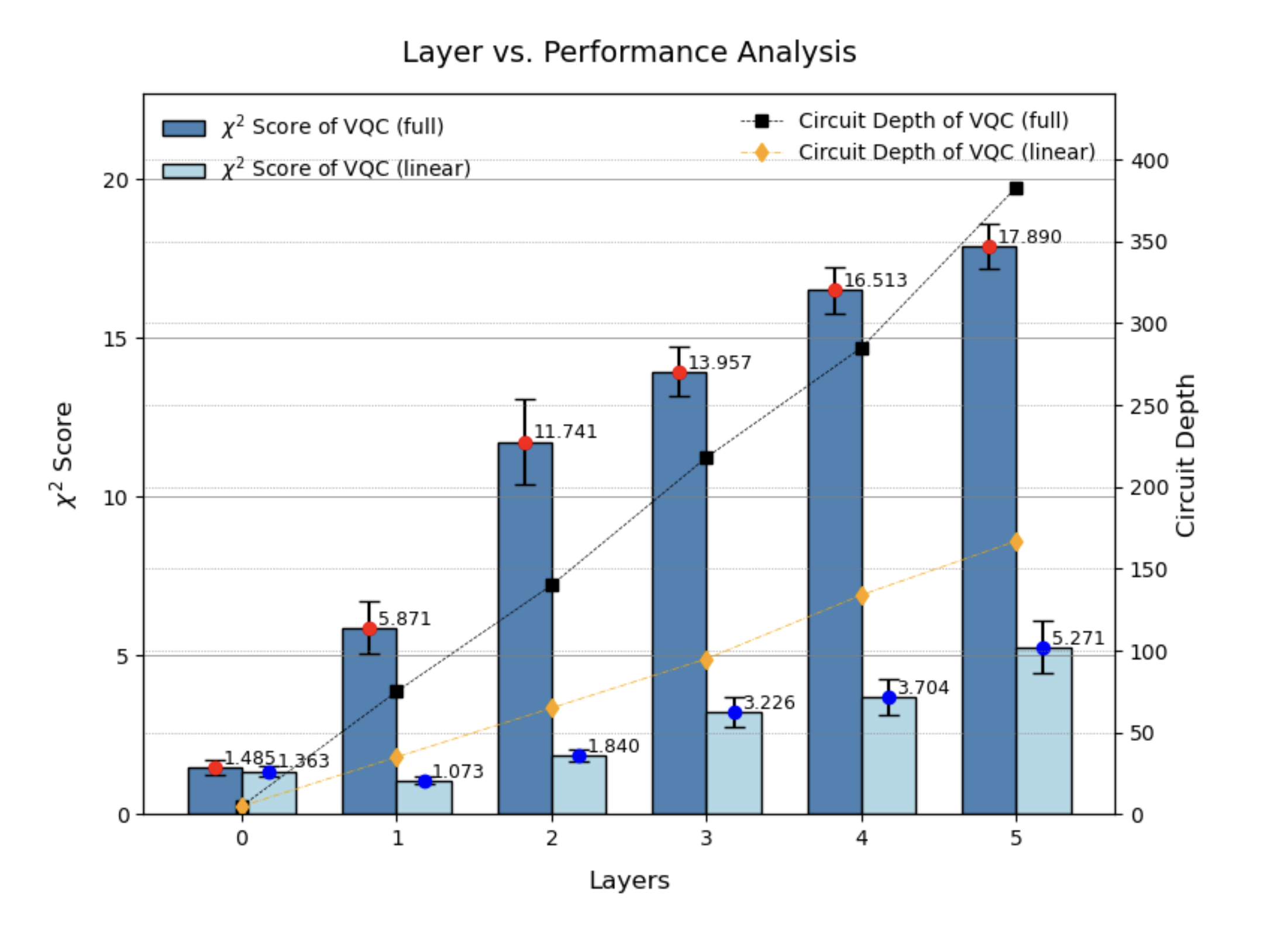}
\caption{$\chi^2$ score and the corresponding circuit depth of full and linear entangler VQC under noisy simulation.}
\label{fig:vqc_circuit_depth}
\end{figure}

\section{Results}
\label{sec:result}
\subsection{Learning QPE under ideal and noisy simulations}
With the data $P^{\text{gt}}(|\phi_i \rangle)$ from Sec.~\ref{sec:quantum_phase_estimation} and the training scheme from Sec.~\ref{sec:vqc_for_QPE}, the results obtained from ideal simulation using $P^{\text{gt}}(|\phi_i \rangle)$ are presented in Fig.~\ref{fig:vqc_ansatz_ideal_result}(c) and Fig.~\ref{fig:vqc_ansatz_ideal_result}(d). The ansatz, indicated as VQC(1, full), denotes $p=1$ with the entangler type ``full" as shown in Fig.~\ref{fig:vqc_ansatz_ideal_result}(b). It is observed that after the training, as evidenced by the declining cost value shown in Fig.~\ref{fig:vqc_ansatz_ideal_result}(d), the resulting measurement in Fig.~\ref{fig:vqc_ansatz_ideal_result}(c) closely resembles that in Fig.~\ref{fig:qpe_result}(a). To quantify the deviation of the measured probability distributions, we employ the $\chi^2$ loss as a metric:
\begin{equation}
\chi^2 = \sum_{i=1}^{2^N} \frac{[P(|\phi_i \rangle) - P^{\text{gt}}(| \phi_i \rangle)]^2}{P^{\text{gt}}(| \phi_i \rangle)},
\end{equation}
where the notation is the same as the cost function in Eq.~(\ref{eq:cost}). A smaller $\chi^2$ value indicates better circuit performance. For instance, the $\chi^2$ value for the result in Fig.~\ref{fig:vqc_ansatz_ideal_result}(c) is $\chi^2 = 0.621$.

\subsubsection{Effect of the VQC layers}

In Fig.~\ref{fig:vqc_layers_entanglers_benchmarking}(a) and Fig.~\ref{fig:vqc_layers_entanglers_benchmarking}(c), the $\chi^2$ results for the full entangler ansatz with varying numbers of layers are displayed under both ideal simulation and noisy simulation conditions. It is evident that, under ideal simulation, as the number of layers increases, the $\chi^2$ value decreases. This phenomenon arises from the circuit becoming more expressive as the number of trainable parameters grows. However, in the case of noisy simulation, the trend of $\chi^2$ is reversed compared to the ideal scenario, as the deeper circuit leads to a significant accumulation of noise.

\subsubsection{Effect of Entangler Kinds}
In Fig.~\ref{fig:vqc_layers_entanglers_benchmarking}(b) and Fig.~\ref{fig:vqc_layers_entanglers_benchmarking}(d), the corresponding results of Fig.~\ref{fig:vqc_layers_entanglers_benchmarking}(a) and Fig.~\ref{fig:vqc_layers_entanglers_benchmarking}(c), but with linear entanglers, are presented. Under ideal simulation, the performance of full entangler and linear entangler showed no statistically significant difference, assured by hypothesis tests. However, under noisy conditions, VQCs with linear entanglers exhibited a significant advantage due to their shallower circuit depth. It was observed that in Fig.~\ref{fig:vqc_layers_entanglers_benchmarking}(d), a single layer of linear entangler, denoted as VQC(1, linear), proved to be the optimal setting. This configuration strikes a balance between the expressibility of the VQC and the accumulation of errors under noisy simulation. In Fig.~\ref{fig:vqc_circuit_depth}, we further investigate the resulting circuit depth of the data in Fig.~\ref{fig:vqc_layers_entanglers_benchmarking}(c) and Fig.~\ref{fig:vqc_layers_entanglers_benchmarking}(d) under noisy simulation. The increasing $\chi^2$ scores for full entanglers are attributed to the accumulating errors arising from deeper circuits. Hence, it is reasonable to employ linear entanglers for experiments conducted under real hardware conditions. Specifically, the setting VQC(1, linear) is used in the following section on real hardware due to its optimality under noisy simulations.

\subsection{Learning QPE on real hardware}
\begin{figure}[ht]
\centering
\includegraphics[scale=0.23]{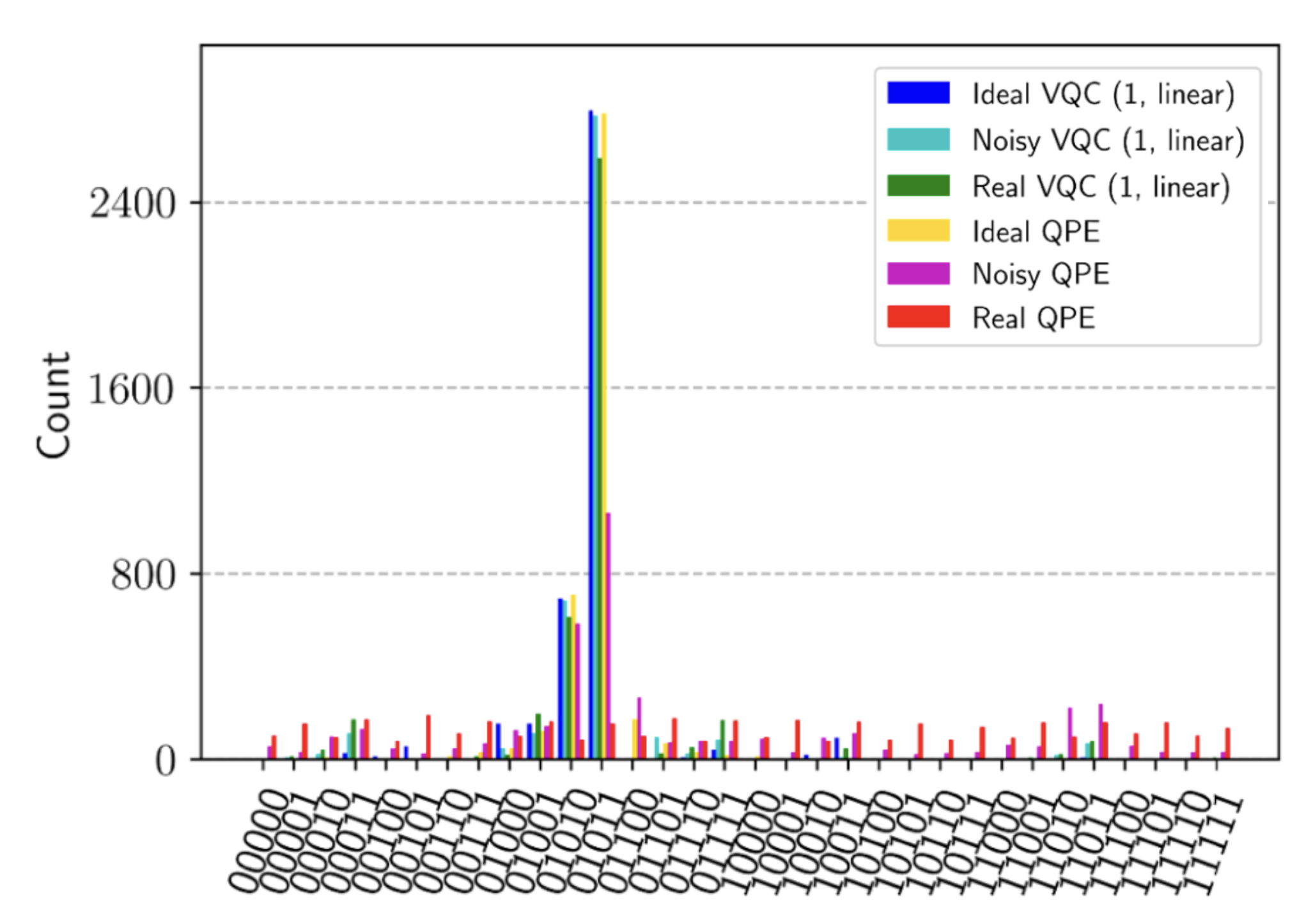}
\caption{Combined measurement results under ideal simulation, noisy simulation, and real hardware.}
\label{fig:vqc_real_hardware_count}
\end{figure}

\begin{figure}[ht]
\centering
\includegraphics[scale=0.21]{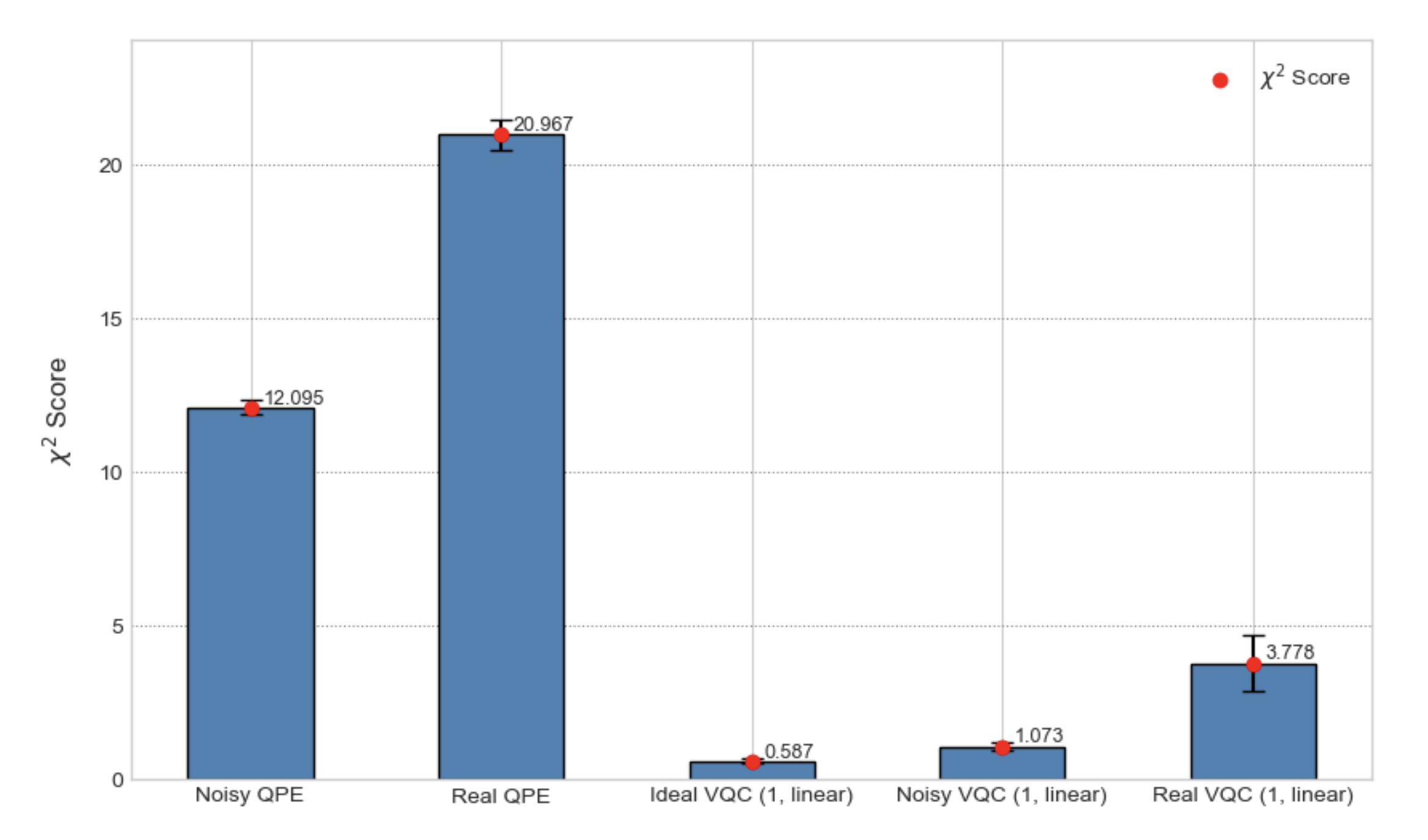}
\caption{The result of the $\chi^2$ scores for different experiment settings.}
\label{fig:vqc_real_hardware_chi}
\end{figure}

The VQC approach significantly reduced the circuit depth. When executed on real hardware, the circuit depth for VQC(1, linear) was only 35, in contrast to the circuit depth of 199 for the standard QPE circuit in Fig.~\ref{fig:qpe_circuit}. Consequently, we expected the performance of VQC (1, linear) to be significantly better compared to the QPE circuit. To assess the performance of VQC(1, linear) on real hardware, we conducted an experiment in which the ansatz parameters were trained under noisy simulations. Subsequently, the trained VQC(1, linear) was executed on the 7-qubit IBM quantum computer \textsf{ibm\_lagos}. In Fig.~\ref{fig:vqc_real_hardware_count}, it is clear that the distributions of VQC(1, linear) closely matched those of ideal QPE, whereas Noisy QPE and Real QPE deviated from the ground truth. To delve into this deviation, the $\chi^2$ scores for the data in Fig.~\ref{fig:vqc_real_hardware_count} are presented in Fig.~\ref{fig:vqc_real_hardware_chi}. These results illustrate that VQC(1, linear) significantly enhanced the performance of QPE under noisy conditions and when executed on real hardware, thus confirming the effectiveness of this approach.

\section{Discussion and Conclusion}
\label{sec:discuss}

The QPE circuits are infamous for their deep circuit depth, which is the main cause of errors. From the experimental results in this work with the standard QPE circuit, we observed significant errors ($\chi^2 = 12.095 \pm 0.243$ in noisy simulation). It performed worse when we executed the same circuits on real hardware ($\chi^2 = 20.967 \pm 0.477$). This demonstrates the need for a shorter circuit to accomplish the same task as QPE, as QPE is often a subunit of a quantum algorithm.


We trained 11 different VQCs (including those with no entangler and 1 to 5 layers of both full and linear entanglers) under both ideal and noisy simulations. In the ideal simulations, both the linear and full entanglers exhibited a trend of decreasing $\chi^2$ scores (the full group's $\chi^2$ score dropped from $0.621\pm0.083$ to $0.195\pm0.032$, and the linear group's $\chi^2$ score dropped from $0.587\pm0.080$ to $0.221\pm0.042")$ as the number of entangler layers increased from 1 to 5. This suggests improved performance resulting from the increased expressibility of additional layers in VQCs. The performance differences between the full and linear entanglers were not evident in the simplicity of the problem under ideal simulation conditions.

However, the disadvantages of the full entangler became apparent and greatly hindered its performance under noisy simulation. The VQC with full entanglers had a significantly deeper circuit, and, as a result, the measurement results were significantly influenced by errors. The $\chi^2$ score increased significantly from $1.485\pm0.238$ to $17.890\pm0.709$ as the number of layers in the full entangler cases increased from 0 to 5. In the case of the linear group, the same trend was observed but in a less severe form. The $\chi^2$ score increased from $1.363\pm0.172$ to $5.271\pm0.831$ as the layers in the linear entangler cases increased from 0 to 5. Interestingly, we observed the expected optimization point where the VQC performed best under noisy simulation.

With the results above, we had VQC(1, linear) trained under noisy conditions and then executed on the real hardware. We found that the VQC(1, linear) had a $\chi^2$ score of $3.778\pm0.899$, which was a significant increase compared to the QPE circuit under the same environment. 

A counterintuitive aspect of this project is the necessity of knowing the ground truth in order to train the VQC. One might wonder why we would go through the trouble of replacing the original circuit with the trained VQC when we already know the phase of QPE. In the case of a standalone QPE, there is indeed no need to train a VQC to replicate the result. However, in practical scenarios, QPE is typically just one component of a larger circuit, and often, it constitutes a significant portion of the circuit's depth. Consequently, in noisy environments, it is necessary to replace the QPE component with the trained VQC to ensure that the overall circuit can proceed with further calculations. With this in mind, we envision our project as a compiler. The compiler's role would involve initially simulating the probability distribution, training the VQC, and subsequently substituting the original circuit with the trained VQC before the circuit is sent for execution. It's important to note that this might require additional steps, such as comprehensive quantum state tomography.

From the insights gained in this project, we have concluded that the linear entangler is a favored choice for NISQ hardware in our particular case. However, this conclusion could be subject to revision when the problem's complexity increases and demands greater expressibility than linear entanglers can provide. In other words, if we intend to replace a different part of the circuit, as opposed to the QPE, we may need a different VQC setting to achieve optimal performance. The systematic analysis of the target circuit and the selection of an appropriate circuit represent future research directions in the context of this ``compiler" approach, aimed at reducing errors in quantum computation.

\section*{Code Availability}
We provide our source codes for project LearnQPE on GitHub: \url{https://github.com/Abeeekoala/LearnQPE}.

\section*{Acknowledgment}
C.Y.L. and C. H. A. L. would like to express their gratitude to Xanadu for hosting QHack2023. This work is an extension of their QHack2023 winning project, LearnQPE. This research received support from the IBM Quantum Researchers Program. K.C. is grateful for the financial support from the Turing Scheme for the Imperial Global Fellows Fund.

\end{document}